\documentclass[10pt]{sigplanconf-pldi16}
\pdfoutput=1 

\usepackage{pervasives}

\begin{document}
\toappear{}

\title{\system: A Configuration Verification Tool for Puppet}

\authorinfo{Rian Shambaugh\quad Aaron Weiss\quad Arjun Guha}{University of Massachusetts Amherst, United States}{\{rian,aaronweiss,arjun\}@cs.umass.edu}

\maketitle

\begin{abstract}
Large-scale data centers and cloud computing have turned system
configuration into a challenging problem. Several widely-publicized
outages have been blamed not on software bugs, but on configuration
bugs. To cope, thousands of organizations use system configuration
languages to manage their computing infrastructure. Of these, Puppet
is the most widely used with thousands of paying customers and many more
open-source users. The heart of Puppet is a domain-specific language
that describes the state of a system. Puppet already performs some basic static
checks, but they only prevent a narrow range of errors. Furthermore,
testing is ineffective because many errors are only triggered under specific
machine states that are difficult to predict and reproduce. With 
several examples, we show that a key problem with Puppet is that
configurations can be non-deterministic.

This paper presents \system, a verification tool for Puppet 
configurations. \system implements a sound, complete, and scalable 
determinacy analysis for Puppet. To develop it,  we (1) present a
formal semantics for Puppet, (2) use several analyses to 
shrink our models to a tractable size, and (3) frame 
determinism-checking as decidable formulas for an SMT solver. \system 
then leverages the determinacy analysis to check other important
properties, such as idempotency. Finally, we apply \system to several
real-world Puppet configurations.
\end{abstract}



\section{Introduction\label{intro}}

Consider the role of a system administrator at any organization, from a large
company to a small computer science department. Their job is to maintain
computing infrastructure for everyone else. When a new
software system, such as a Web service, needs to be deployed, it is their job to
provision new servers, configure the firewall, and ensure that data is
automatically backed up. If the Web service receives a sudden spike in traffic,
they must quickly deploy additional machines to handle the load.  When a
security vulnerability is disclosed, they must patch and restart machines if
necessary. All these tasks require the administrator to write and maintain
system configurations.

Not too long ago, it was feasible to manage systems by directly
running installers, editing configuration files, etc. A skilled administrator
could even write shell scripts to automate some of these tasks.
However, the scale of modern data centers and
cloud computing environments has made these old approaches brittle
and ineffective.

%

\paragraph{System configuration languages.}
System configuration is a problem that naturally lends itself to domain-specific
languages (DSLs). In fact, the programming languages community has developed
several DSLs for specifying system configurations that are
used in practice. For example, NixOS~\cite{dolstra:nixos} uses a lazy, functional language to describe
packages and system configurations;  Augeas~\cite{augeas} uses
lenses~\cite{bohannon:string-lenses} to update
configuration files; and
Engage~\cite{fischer:engage} provides a declarative DSL that tackles
issues such as inter-machine dependencies.

In the past few years, several system configuration languages have also
been developed in industry. Puppet, Chef, and Ansible (recently
acquired by Red Hat) are three prominent examples. This paper focuses
on Puppet, which is the most popular of these languages, but
these commercial languages have several features in common that set 
them apart
from prior research. First, they support a variety of operating
systems, tools, and techniques that systems administrators already know. Unlike
NixOS, they don't posit new package managers or new Linux distributions, but
simply use tools like \texttt{apt} and \texttt{rpm} under the hood. Second, these languages
provide abstractions for managing several kinds of resources, such as packages,
configuration files, user accounts, and more. Therefore, they are broader in
scope than Augeas, which only edits configuration files. Finally, these
languages provide relatively low-level abstractions, compared to earlier work
like LCFG~\cite{anderson:lcfg}.
For example, Puppet provides a simple and expressive DSLs that encourages average
users to build their own abstractions.

\paragraph{Puppet.}
Puppet configurations (called \emph{manifests})
are written in a expressive, yet constrained DSL, which makes them
amenable to analysis.
To a first approximation, a manifest
specifies a collection of resources, their desired state, and their
inter-dependencies. For example, the following Puppet manifest states
that the \verb|vim| package should be installed, that the user account 
\verb|carol| should exist, and that she should have a \verb|.vimrc| 
file in her home directory containing the single line 
\texttt{syntax on}:
\begin{puppetcode}
package{'vim': ensure => present }
file{'/home/carol/.vimrc': content => 'syntax on' }
user{'carol': ensure => present; managehome => true }
\end{puppetcode}
It's tedious to  describe every individual resource in this
manner, so Puppet makes it easy to write parameterized modules.
The official module repository, Puppet Forge, has
nearly four thousand modules from over six hundred contributors.

\paragraph{Non-determinism and modularity.}
A key property of Puppet is that manifests should be 
deterministic~\cite{puppet-determinism}.
Determinism is a critical property because it helps ensure that
a manifest has the same effect in testing and in production.
Similarly, if one manifest is applied to several machines, which is 
common in large deployments, determinism helps to ensure that they are 
replicas of each other.

Unfortunately, it is  easy to write manifests that are not deterministic.
Puppet can  install resources in any order, unless the manifest explicitly
states inter-resource dependencies.\footnote{Puppet calculates
dependencies automatically only in some trivial cases, \eg, files
``auto-require'' their parent directory.}
Therefore, the example manifest above
is non-deterministic: there will be a runtime error  if Puppet tries to create the
file \verb|/home/carol/.vimrc| before Carol's account.
We can fix this bug by making the dependency explicit:
\begin{puppetcode}
User['carol'] -> File['/home/carol/.vimrc']
\end{puppetcode}

The fundamental problem is that Puppet manifests specify a partial-order on
resources, thus resources can be installed in several orders. However, when some dependencies are missing, applying the
manifest can go wrong: the system may signal an error or may even fail silently
by transitioning to an unexpected state. These bugs are very hard to detect with
testing, since the number of valid permutations of resources becomes 
intractable very quickly.

Surprisingly, a manifest can also have too many
dependencies and be over-constrained.
Imagine two manifests $A$ and $B$ that
both install the resources $R_1$ and $R_2$.
Suppose that $R_1$ and $R_2$ do not depend on each other, but the
manifest authors take a conservative approach and add a false dependency to
avoid non-determinism issues. If
$A$ picks $R_1~\texttt{->}~R_2$ and $B$ picks $R_2~\texttt{->}~R_1$ then
$A$ and $B$ cannot be composed.

Therefore, manifests must be deterministic to be correct,  but must only have
essential dependencies to be composable. Without composability, manifests cannot
be decomposed into reusable modules, which is one of the key features of Puppet.
However, when a manifest is only partially-ordered, we may need to check an
intractably large number of orderings to verify determinism.

A further complication is that the Puppet has a diverse collection
of resource types, which makes it hard to determine how resources 
interact with each other. For example, a file may overwrite another
file created by a package,
a user account may need the \verb|/home| directory to be present,
a running service may need a package to be installed, and so on.
We could try to side-step this issue by building a dynamic determinacy
analysis~\cite{burnim:determinism,sadowski:singletrack}. However, a purely dynamic
approach could only identify a problems when two replicas
diverge, whereas a static determinacy analysis helps ensure that a
manifest behaves correctly on any machine regardless of its initial
state.

\paragraph{Idempotency.}
Determinism is not a sufficient condition to ensure that
Puppet behaves predictably. In a typical deployment, the Puppet background
process periodically reapplies the manifest to ensure that the
machine state is consistent with it. For example, if a user modifies
the machine (\eg, manually editing configurations), re-applying the
manifest will correct the discrepancy. Thus if the machine state has not
changed, reapplying the manifest should have no
effect.
Like determinacy, this form
of idempotence is also believed to be a key property of Puppet~\cite{puppet-idempotence}. However, it is
also trivial to construct manifests that are not idempotent.

\paragraph{Our approach.}
To the best of our knowledge, this is the first paper to develop 
programming language techniques for Puppet (or a related language such
 as Chef and Ansible). We first present a core fragment of Puppet with
several small examples that illustrate its problems
(\cref{overview}). We develop a formal semantics of Puppet that models
manifests as programs in a simple, non-deterministic imperative 
language of filesystem operations called \corelang (\cref{semantics}).

Our main technical result is a sound, complete, and scalable determinacy
analysis (\cref{determinacy}). To scale to real-world examples, we use
three different analyses to shrink the size of models. The first
two analyses dramatically reduce the number of paths that the determinism-checker
needs to reason about by eliminating resources that do not affect determinism and eliminating other side-effects that
are not observed by the rest of the program. The third analysis is an unusual commutativity check
that accounts for the fact that resources are mostly idempotent. Finally
after leveraging the aforementioned analyses, the determinacy checker encodes the
semantics as effectively-propositional formulas for an SMT solver.

We argue that our determinacy analysis enables several other higher-level
properties to be checked (\cref{idempotence}), and show this is the case by developing
a simple idempotence checker that leverages determinism in a fundamental way.

We implement our algorithms in a tool called \system, which we evaluate on
several real-world examples (\cref{evaluation}). Finally, we discuss related work (\cref{related}), summarize the limitations of our
approach (\cref{limitations}), and conclude (\cref{conclusion}). The \system source code, benchmarking scripts, and a technical appendix are available online~\cite{guha:rehearsal-tool}.

\section{Introduction to Puppet\label{overview}}

\begin{figure}
\figsize
\begin{minipage}{.45 \textwidth}
\(
\begin{array}{@{}r@{~~}r@{~}c@{~}l@{~~}l@{}}
\textrm{Types} & \textit{rtype} & \Coloneqq & \texttt{file} \mid \texttt{package} \mid \cdots \\
\textrm{Strings} & \textit{str} & \Coloneqq & \texttt{"$\cdots$\$x$\cdots$\$y$\cdots$"} \\
\textrm{Identifiers} & x & \Coloneqq & \texttt{\$x} \mid \texttt{\$y} \mid \cdots \\
\textrm{Titles} & t & \Coloneqq & \textit{str} \mid x \\
\textrm{Values} & v & \Coloneqq & \textit{str} & \textrm{String} \\
                          &   & \mid      & n            & \textrm{Number} \\
                          &   & \mid      & \texttt{[} v_1 \cdots v_n  \texttt{]} & \textrm{Array} \\
                          &   & \mid      & x            & \textrm{Variable} \\
\textrm{Attributes} & \textrm{attr} & \Coloneqq & \textit{str}~\texttt{=>}~v \\
\textrm{Resources}
& R & \Coloneqq
& \textit{rtype}\texttt{\{}t\texttt{:\ }\textit{attr}_1\cdots\textit{attr}_n \texttt{\}} \\
\textrm{Manifests}
& m & \Coloneqq
  & R
  & \textrm{Resource} \\
& & \mid
  & \texttt{define}~\textit{rtype}\texttt{(}x_1 \cdots x_n \texttt{) \{}~m~\texttt{\}}
  & \textrm{Type} \\
& & \mid
  & \textit{rtype}_1\texttt{[}t_1\texttt{]}~\texttt{->}~\textit{rtype}_2\texttt{[}t_2\texttt{]}
  & \textrm{Dependency} \\
& & \mid
  & m_1~m_2
  & \textrm{Composition}
\end{array}
\)
\end{minipage}
\caption{Syntax of Puppet fragment used in this paper.}
\label{corepuppet}
\end{figure}



\begin{figure}
\begin{puppetcode}
define myuser($title) {
  user {"$title":
    ensure => present,
    managehome => true
  }
  file {"/home/${title}/.vimrc":
    content => "syntax on"
  }
  User["$title"] -> File["/home/${title}/.vimrc"]
}

myuser {"alice": }
myuser {"carol": }
\end{puppetcode}
\caption{A user-defined resource type and its instantiations.}
\label{myuser}
\end{figure}

This section introduces the fragment of Puppet that we use in the
exposition of this paper. We also illustrate the kinds of problems
that \system solves.

\subsection{A Core Fragment of Puppet}

The Puppet DSL is quite sophisticated. It has typical features such as
functions, loops, and conditionals, and several domain-specific features
that make it easy to specify resources and their relationships.
\system can parse and process a significant subset of Puppet, but, for
clarity, we constrain our examples to the fragment of Puppet
shown in \cref{corepuppet}. A manifest, $m$, is composed of
resources,
resource type declarations, and inter-resource dependencies.
A resource, $R$, has a type, a title, and a map of attributes.
The resource type determines how the attribute-map is interpreted.
For example, a \verb|file| resource must have an attribute called \verb|path|,
a \verb|user| resource must have an attribute called \verb|name|, and so on.
The resource title can be any descriptive string, but is often used as the default
value for an essential attribute. For example, if a \verb|file| resource does
not have the required \verb|path| attribute, the title is used as the \verb|path|.
A manifest can declare several resources by juxtaposition, but the order
in which resources appear is not significant. Instead, manifests must specify
dependencies explicitly. To state that the resource $t_2$ depends
on the resource $t_1$, we write
$\textit{rtype}_1\texttt{[}t_1\texttt{]}~\texttt{->}~\textit{rtype}_2\texttt{[}t_2\texttt{]}$.\footnote{The first letter of a type name is capitalized in resource 
references.}
In addition to a few dozen built-in resource types, Puppet allows manifests
to define their own types. A type definition is essentially a function
that consumes named attributes as arguments and produces a manifest as a result.
For example, if all users in an organization use the same default environment,
we can create a new type called \verb|myuser| and instantiate it for several
users, as shown in \cref{myuser}.

\subsection{Common Puppet Problems}

There are a number of problems that can easily occur in Puppet manifests.

\begin{figure}
\begin{minipage}{\columnwidth}
\begin{puppetcode}
file {"/etc/apache2/sites-available/000-default.conf":
  content => ...,
}

package{"apache2": ensure => present }
\end{puppetcode}
\subcaption{Signals an error nondeterministically.}\label{apache2}
\end{minipage}
\vskip .5em
\begin{minipage}{\columnwidth}
\begin{puppetcode}
define cpp() {
  package{'m4': ensure => present }
  package{'make': ensure => present }
  package{'gcc': ensure => present }
  Package['m4'] -> Package['make']
  Package['make'] -> Package['gcc']
}

define ocaml() {
  package{'make': ensure => present }
  package{'m4': ensure => present }
  package{'ocaml': ensure => present }
  Package['make'] -> Package['m4']
  Package['m4'] -> Package['ocaml']
}
\end{puppetcode}
\subcaption{Cannot be composed due to false dependencies.}\label{cpptype}
\end{minipage}
\vskip .5em
\begin{minipage}{\columnwidth}
\begin{puppetcode}
package{'golang-go': ensure => present }
package{'perl': ensure => absent }
\end{puppetcode}
\subcaption{Leads to two different success states.}
\label{pp_flipflop}
\end{minipage}
\vskip .5em
\begin{minipage}{\columnwidth}
\begin{puppetcode}
file{"/dst": source => "/src" }
file{"/src": ensure => absent }
File["/dst"] -> File["/src"]
\end{puppetcode}
\subcaption{Not idempotent.}\label{pp_flipflop2}
\end{minipage}

\caption{Several problematic manifests.}
\end{figure}

\paragraph{Non-deterministic errors.}
A common Puppet idiom is to first install a package and then overwrite its
default configuration. For example, the \verb|apache2| package installs
a web server and several configuration files.
To host a website, at least the default site configuration file,
\verb|000-default.conf|, has to be replaced (\cref{apache2}).
If the dependency between the package and the file is accidentally omitted,
Puppet may try to create the configuration file first which would signal
an error because the file is in a directory that the package has yet to create.

\paragraph{Over-constrained dependencies.}
Consider a strawman solution to the non-determinism problem: we could add false
dependencies so that all resources are totally ordered. Unfortunately,
this approach makes it difficult to write independent modules which is
one of the main features of Puppet. For example, \cref{cpptype} shows
two simple types that configure C++ and OCaml development
environments.\footnote{Idiomatic Puppet would use
the \texttt{class} keyword.} Both modules install \texttt{make}
and \texttt{m4} because they are commonly used by C++ and OCaml projects.
To force determinism, both modules in the figure have false dependencies between
\texttt{make} and \texttt{m4}. However, each has picked a different order
 which can easily occur when the modules have different authors.
Therefore, if we try to instantiate both modules simultaneously, Puppet will fail
and report a dependency cycle.~\footnote{Readers familiar with Puppet may know
that shared resources have to be guarded with \texttt{defined}. Some
people consider \texttt{defined} to be an anti-pattern, but a simple search
shows that it is used in over 1/3rd of the packages on Puppet Forge
to enable the kind of modularity that we discuss.}
This heavy-handed approach to determinism sacrifices composability.

\paragraph{Silent failure.}
In addition to non-deterministic errors, it is also possible to write a manifest that
nondeterministically leads to two distinct states without Puppet reporting an error.
For example, the manifest in \cref{pp_flipflop} states that Perl should be
removed and the Go compiler should be installed. Surprisingly, on Ubuntu 14.04, 
the Go compiler
depends on Perl\cite{go-depends-on-perl}, so this state is not realizable. Puppet cannot detect this
problem, but simply dispatches to the native package manager (\eg,
\verb|apt| or \verb|yum|) to actually install and remove packages. For this
manifest, Puppet
issues two low-level commands to remove Perl and install Go. Since there are no
dependencies, they may execute in either order. If Perl is first removed, the
command to install Go installs Perl too, but if Perl is removed after Go is
installed, that command will remove Go too. This kind of error is more
insidious than a nondeterministic error, since there isn't an obvious fix.

\paragraph{Non-idempotence.}
Another key property of Puppet manifests is that they should be idempotent:
applying a manifest twice should be the same as applying it once. However,
Puppet does not enforce this property, which makes it easy to produce manifests that
are not idempotent. For example, we can make the non-deterministic manifest
in \cref{pp_flipflop} deterministic by removing Perl before Go
is installed:
\begin{puppetcode}
Package['perl'] -> Package['golang-go']
\end{puppetcode}
However, this manifest is not idempotent. Puppet checks which packages
are installed before it issues any commands to install or remove packages.
In this example, if both packages are already installed, Puppet will
remove Perl and take no further action, even though removing
Perl removes Go. If we apply the manifest again (\ie, when neither
package is installed), Puppet installs Go
and takes no further action, even though Perl is implicitly installed.
The real issue is that this manifest is fundamentally inconsistent and
cannot be fixed by adding dependencies. A system cannot have Perl
removed and Go installed, so the manifest should be rejected.

An even simpler example of non-idempotence is the manifest in
\cref{pp_flipflop2}, which copies  \verb|src| to \verb|dst| and
then deletes \verb|src|. The second run of this manifest will always fail,
because the first run removes
\verb|src|. This example shows that even though primitive resources
are designed to be idempotent, they can be composed in ways that break
idempotence.

\paragraph{Summary.}
We've introduced a small fragment of Puppet and used it to illustrate several kinds
of bugs that can occur in Puppet manifests. We've argued that the root cause of these
bugs is that Puppet does not ensure that manifests are deterministic and idempotent.
Before we describe how \system{} checks these properties, we present the semantics of
Puppet that \system{} uses.

\section{Semantics of Puppet\label{semantics}}

This section presents a semantics for Puppet, which we develop in two stages.
(1) We compile manifests to a directed acyclic graph of primitive resources,
which we call a \emph{resource graph}. The compilation process involves several
passes to eliminate features that inject dependencies, change attributes, and so
on. We also substitute instantiations of user-defined types with their
constituent resources until only primitive resources remain. (2) Next, we model the semantics of individual resources as programs in a
small imperative language of file system operations called \corelang.  We
carefully design \corelang so that it is expressive enough to describe the
semantics of resources, yet restrictive enough to enable the static analyses we
present in subsequent sections.

\subsection{From Puppet to Resource Graphs\label{puppet2rg}}

\begin{figure}
\figsize
\(
\begin{array}{lrcl}
\textrm{Vertices}
& V & \Coloneqq & v_1 \mid \cdots \mid v_k  \\
\textrm{Edges}
& E & \subseteq & V \times V \\
\textrm{Vertex Labels}
& L & \in & V \rightarrow R \\
\textrm{Resource Graphs}
& G & \in & V \times E \times L
\end{array}
\)
\caption{Resource graphs.}
\end{figure}

A \emph{resource graph} $G$ is a directed acyclic graph with vertices labeled by
primitive resources. An edge exists from $V_1$ to $V_2$ if $V_2$ depends on
$V_1$. At a high-level, we compile manifests into resource graphs by converting
primitive resources to nodes and dependencies to edges. To do this, we
employ a number of passes to simplify manifests.

Puppet has several abstractions that allow manifests to succinctly describe dependencies.
For example, user-defined types can be used to abstract over a collection of other resources and dependencies.
We reduce user-defined abstractions to their constituent
resources by repeatedly substituting their definitions until only primitive resources
remain. In order to preserve ordering, this pass must introduce new edges
between resources within instances of abstractions. In addition, resources can also
be assigned to a \emph{stage}, and stages are ordered independently of resources. To
deal with this, we implement a stage elimination pass that adds edges between the
constituent resources of each stage.

Certain Puppet features have non-local side effects. For example, the
following expression uses a  \emph{resource collector} to update all
file-resources owned by \verb|carol| to be unreadable by others,
regardless of where they are defined:
\begin{puppetcode}
File<| owner == 'carol' |> { mode => "go-rwx" }
\end{puppetcode}
Unfortunately, resource collectors are not modular and make separate
compilation impossible. In general, it is not possible to know the
attributes of a resource until all user-defined types
(which may define collectors) are eliminated as described above.
The passes that tackle these kinds of expressions are necessarily global
transformations.

Our compiler tackles the details described above and some other features
of Puppet that we don't belabor here.

\subsection{From Resources to \corelang Programs}

\begin{figure*}
\figsize
\begin{minipage}{.384\textwidth}
\(
\begin{array}{@{}r@{~~}r@{~}c@{~}l@{~}l@{}}
\multicolumn{4}{@{}l}{\textbf{Syntax}}\\[.25em]
\textrm{Paths} & p &
  \Coloneqq & \texttt{/}                            & \textit{Root directory} \\
   & & \mid & p \texttt{/} \textit{str}            & \textit{Sub-path} \\[.25em]
\textrm{File Contents} & v &
  \Coloneqq & \dir & \textit{Directory} \\
   & & \mid & \file{\textit{str}} & \textit{File} \\[.25em]
\textrm{File Systems} & \fs &
  \Coloneqq & \multicolumn{2}{l}{\langle p_1\!= \! v_1 \cdots p_k \! = \! v_k \rangle} \\[.25em]
\textrm{Predicates} & a &
  \Coloneqq & \pstat{p}{\dne}                     & \textit{Does not exist} \\
   & & \mid & \pstat{p}{\isfile}                  & \textit{Is a file} \\
   & & \mid & \pstat{p}{\isdir}                   & \textit{Is a directory} \\
   & & \mid & \pstat{p}{\isempty}                 & \textit{Is an empty dir.} \\
   & & \mid & \ptrue                              & \textit{True} \\
   & & \mid & \pfalse                             & \textit{False} \\
   & & \mid & \por{a_1}{a_2}                      & \textit{Disjunction} \\
   & & \mid & \pand{a_1}{a_2}                     & \textit{Conjunction} \\
   & & \mid & \pnot{a}                            & \textit{Negation} \\[.25em]
\textrm{Expressions} & e &
  \Coloneqq & \eskip                        & \textit{No op} \\
   & & \mid & \eerror                       & \textit{Halt with error} \\
   & & \mid & \mkdir{p}                     & \textit{Create directory} \\
   & & \mid & \mkfile{p}{\textit{str}}      & \textit{Create file} \\
   & & \mid & \rmfile{p}                    & \textit{Remove file/empty dir.} \\
   & & \mid & \cp{p_1}{p_2}                 & \textit{Copy file} \\
   & & \mid & \seq{e_1}{e_2}                & \textit{Sequencing} \\
   & & \mid & \ite{a}{e_1}{e_2}             & \textit{Conditional}
\end{array}
\)
\end{minipage}
\vrule
\figsize
\begin{minipage}{.6 \textwidth}
\noindent \(
\begin{array}{r@{~}c@{~}l}
\multicolumn{3}{@{\quad}l}{\textbf{Semantics}}\\[.25em]
\den{a} & \in & \fs \rightarrow \kw{bool} \\[.25em]
\den{\pstat{p}{\isfile}}\fs & \defeq & \exists \textit{str}. \fs(p) = \file{\textit{str}} \\
\den{\pstat{p}{\isdir}}\fs & \defeq & \fs(p) = \dir \\
\den{\pstat{p}{\dne}}\fs & \defeq & p \not\in \textit{dom}(\fs)  \\
\den{\pstat{p}{\isempty}}\fs & \defeq &
  \fs(p) = \dir~\textrm{and}~\neg\exists \str . \mkpath{p}{\str} \in \textit{dom}(\fs) \\
\cdots
\\[.25em]
\den{e}      & \in & \fs \rightarrow \fs + \errst \\[.25em]
\den{\eskip} \fs & \defeq & \fs \\
\den{\eerror} \fs & \defeq & \errst \\
\den{\mkdir{\mkpath{p}{\str}}} \fs & \defeq &
  \left\{\begin{array}{@{}ll}
    \updfs{\fs}{\mkpath{p}{\str}}{\dir} &
      \den{\pand{\pstat{p}{\isdir}}{\pstat{\mkpath{p}{\str}}{\dne}}}\sigma \\
    \errst &
      \textrm{otherwise}
  \end{array}\right. \\[.25em]
\den{\mkfile{\mkpath{p}{\str}}{\textit{str}'}} \fs & \defeq &
  \left\{\begin{array}{@{}ll}
    \updfs{\fs}{\mkpath{p}{\str}}{\file{\textit{str}'}} &
      \den{\pand{\pstat{p}{\isdir}}{\pstat{\mkpath{p}{\str}}{\dne}}}\fs \\
    \errst &
      \textrm{otherwise}
  \end{array}\right. \\
\den{\rmfile{p}} \fs & \defeq &
  \left\{\begin{array}{@{}ll}
    \fs - p &
      \den{\por{\pstat{p}{\isfile}}{\pstat{p}{\isempty}}}\fs \\
    \errst &
      \textrm{otherwise}
  \end{array}\right. \\
\den{\cp{p_1}{\mkpath{p_2}{\str}}} \fs & \defeq &
  \left\{\begin{array}{@{}ll}
    \updfs{\fs}{\mkpath{p_2}{\str}}{\file{\str'}} &
      \den{\pand{\pstat{\mkpath{p_2}{\str}}{\dne}}{\pstat{p_2}{\isdir}}}\fs \\
      & \textrm{and}~\fs(p_1) = \file{\str'} \\
    \errst &
      \textrm{otherwise}
  \end{array}\right. \\
\den{\seq{e_1}{e_2}} \fs & \defeq &
  \left\{\begin{array}{@{}ll}
    \den{e_2}\fs' & \den{e_1}\fs = \fs' \\
    \errst & \den{e_1}\fs = \errst
  \end{array}\right. \\
\den{\ite{a}{e_1}{e_2}} \fs & \defeq &
  \left\{\begin{array}{@{}ll}
    \den{e_1} \fs & \den{a}\fs \\
    \den{e_2} \fs & \textrm{otherwise}
  \end{array}\right.
\end{array}
\)
\end{minipage}
\caption{\corelang syntax and semantics.}
\label{corelang-fig}
\end{figure*}

Puppet has dozens of different primitive resource types that can interact
with each other in subtle ways. Moreover, some
resources have flags that dramatically change their behavior. To deal
with this diversity, we model resources as small programs in a low-level
language called \corelang that captures their essential effects and possible interactions.
The advantage of using \corelang is that we can quickly add support for
additional resource types and new versions of Puppet without rebuilding the
rest of our analysis toolchain. In this paper, \corelang is an
imperative language with simple operations that affect the filesystem.
However, it also would be straightforward to enrich the language in several ways.

\paragraph{Syntax and semantics of \corelang.}
The \corelang language, defined in \cref{corelang-fig} is a simple imperative
language of programs that manipulate the filesystem. We model filesystems (\fs)
as maps
from paths ($p$) to file contents. A file may be a regular file with some
content (\file{\str}) or the value \dir{} that represents a directory.
Expressions in \corelang denote functions that consume filesystems and produce
either a new filesystem or error ($\errst$). \corelang has primitive expressions
to create directories (\mkdir{p}), create files (\mkfile{p}{\str}), remove files and
empty directories (\rmfile{p}), and copy files (\cp{p_1}{p_2}).
Sequencing (\seq{e_1}{e_2}) and conditionals (\ite{a}{e_1}{e_2}) behave
in the usual way. Predicates include the usual boolean connectives and primitive
 tests to check if a path is a  file (\pstat{p}{\isfile}),
 a directory (\pstat{p}{\isdir}), an empty directory (\pstat{p}{\isempty}), 
 or contains nothing (\pstat{p}{\dne}).

Since \corelang has no loops, its programs always terminate. This is a
reasonable restriction since applying Puppet resources must terminate
too. \corelang does not have procedures or variables, but their omission
doesn't affect programmers, since \corelang code is generated from
a host language (in our case, Scala). A more important restriction is that
 \corelang programs work with a
finite set of paths and file contents, so \corelang programs are finite state. At first glance, it appears that a program would not be
affected by the state
of paths that do not appear in the program text. However, the
semantics of $\rmfile{p}$ and $\pstat{p}{\isempty}$ are affected by
subpaths of $p$, even if they don't appear in the program.
Finally, \corelang programs only work with a finite set of file
contents. In fact, there are no operations that allow programs to read the
contents of files, but this is not an essential property.

\corelang can easily be extended in several ways to produce higher-fidelity
models of Puppet resources. \eg, it is easy to imagine
adding timestamps, file-permissions, and so on. Notably, these extensions
would not affect the finiteness of \corelang programs, so we believe our
analysis approach would work with these higher-fidelity models too.

\paragraph{Notation.} We write $e_1 \deneq e_2$ when both expressions produce
the same output (or error) for all input filesystems.  For brevity,
we use \ifthen{e_1}{e_2} as shorthand for \ite{e_1}{e_2}{\eskip}.

\subsection{Modeling Resources as \corelang Programs\label{compilingresources}}

Now that we have a language of filesystem operations,  we define a
compilation function $\mathcal{C}: R \rightarrow e$ that maps resources to \corelang
expressions. The actual definition has several hundred lines of code and is
quite involved, but the high-level idea is to model each
resource as an \corelang{} program. Even for simple resources, $\mathcal{C}$ needs to validate
attributes, fill-in values for optional attributes, and produce programs that
check several preconditions before applying the desired action.
We now illustrate how $\mathcal{C}$ models several key resource-types.

\paragraph{Files and directories.}
Individual files and directories are
the simplest resource in Puppet. The \verb|file| resource type manages both and
has several attributes that determine (1) whether it is a file or directory, (2)
if it should be created or deleted, (3) if parent directories should be created,
(4) the contents of a file, or (5) a source file that is copied over. Moreover,
all combinations of these attributes are not meaningful, and most are optional.
The $\mathcal{C}$ function addresses these details in full.

\paragraph{SSH keys.}

Some Puppet resources edit the contents of configuration files. For example,
the \verb|ssh_authorized_key| resource manages a user's public keys, where
each resource is an individual line of a single file. Rather than increase the
complexity of \corelang by including detailed file-editing commands, we model
the logical structure of these resources in a portion of the filesystem disjoint
from other files. However, this alone disguises a certain kind of determinacy bug.
Consider a manifest with two resources: an \verb|ssh_authorized_key| and a \verb|file|
that overwrites the key-file. Clearly, these resources do not commute, but by placing
ssh keys in their own disjoint directory, the compiled program would be deterministic.
To address this issue, our model for \verb|ssh_authorized_key| also creates a key-file
and sets its content to a unique value, enabling us to catch this kind of determinacy bug.

\paragraph{Packages.} A package resource creates (or removes)
a large number of files and directories, so we need this file list
to model packages. Fortunately, there are simple command-line tools that do exactly this:
\eg, \verb|apt-file| for Debian-based systems, \verb|repoquery| for Red Hat-based systems,
and \verb|pkgutil| for Mac OS X.\footnote{We've tested with \texttt{apt-file}
and \texttt{repoquery}.} The
$\mathcal{C}$ function invokes the aforementioned tool and builds a (potentially
very large) program that first creates the directory tree and then issues a
sequence of $\mkfile{p}{\str}$ commands to create the files.  In our model, we
simply give every file $p$ in a package a unique content $\str$.
 This model is sound but conservative: some equivalences can be lost. For
example, suppose a manifest has two resources: a package that creates a file $p$
and a file resource that overwrites $p$ with exactly the same contents as the
package. This manifest would be deterministic without any dependencies, but our
tool would report it as nondeterministic, due to our conservative package
model. However, this situation is unlikely to arise in practice, but if it does
it may indicate another mistake: it's more likely that the author meant
to overwrite $p$ with some other contents.

\paragraph{Other resource types.}
We model several other resource types, including cron jobs, users,
groups,  services, and host-file entries. Puppet has
several resources types that are only applicable to Mac OS X or Windows
systems that we have not modeled. However,
if we wished to analyze a manifest for these platforms, it should be easy
to extend the resource compiler to support these resources. Notably, the
rest of our toolchain would be unchanged as it is agnostic to the actual
set of resources since it operates over \corelang programs.

\begin{figure}
\figsize
\(
\begin{array}{r@{~}c@{~}l}
\den{G} & \in & \fs \rightarrow 2^{\fs + \errst} \\
\den{G} \fs & \defeq & \{ \den{\mathcal{C}(L(v_1)); \cdots; \mathcal{C}(L(v_n))}\fs
\mid \langle v_1\cdots v_n\rangle \in \kw{perms}(G) \} \\
& & \qquad\textrm{where}~G=(V,E,L)
\end{array}
\)
\caption{Semantics of resource graphs.}
\label{rgsemantics}
\end{figure}

\subsection{Semantics of Resource Graphs\label{compilerfinal}}

Now that we have a compiler from resources to \corelang, it is straightforward
to give a semantics to resource graphs. Informally, a resource graph denotes a function from
filesystems to a set of filesystems and the error state. To define this
function, we take all sequences of resources that respect the order imposed by
the edges, compile each resource-sequence to a sequence of \corelang programs, apply
each program to the input state, and take the union of the results
(\cref{rgsemantics}).

A pleasant feature of this definition is that the resource graph and resource
compiler abstract away the peculiarities of Puppet. We can extend $\mathcal{C}$
to support new resource types or the Puppet compiler to support even more Puppet
features without changing the methods that will be discussed in the rest of this paper.

\section{Determinacy Analysis\label{determinacy}}

This section presents our main technical result
which is a sound, complete, and scalable approach to check that
resource graphs (produced from manifests) are deterministic.
\begin{definition}[Determinism]
A resource graph $G$ is deterministic, if for all filesystems $\fs$,
$\left\vert\den{G}\fs\right\vert = 1$.
\end{definition}
This property does not preclude a manifest from always producing an
error on some or even all inputs. Any non-trivial manifest makes
assumptions about the initial state (\eg, the operating system in use) and thus will raise an error if it is applied to a machine that is not in the right initial state. Determinism simply guarantees that successes and failures will be predictable.

Our approach has three major steps:
\begin{enumerate}

\item The first step is to reduce the number of paths that we need to reason
about. Even a small manifest may manipulate several hundred
paths and tracking their state over hundreds of intermediate states can be
intractable. We observe that resources often modify paths $p$
that are not accessed by any other resource, thus operations on these paths
can be safely eliminated without affecting the result of the determinism-check.

\item The next step is to reduce the number of permutations of the resource
graph, which can grow exponentially with the number of resources.
The natural approach is to use partial-order
reduction with a fast, commutativity check. However, the obvious
approach, based on calculating read- and write-sets is not effective
because many resources may create overlapping directories (\eg, \texttt{/usr} and
\texttt{/etc}). We observe that this is a form of false sharing and
develop a commutativity check that accounts for idempotent directory creation.

\item The final step is to encode the semantics of the manifest as a decidable formula for
an SMT solver that is satisfiable if and only if the program is
non-deterministic. Our encoding relies on the fact that programs manipulate a
finite set of paths that are statically known. However, the result of some
operations may be affected by the state of paths that do not appear in the
program itself. We carefully bound the domain of paths to ensure our approach is
complete.

\end{enumerate}
We first present our encoding of manifests as formulas.

\begin{figure*}
\figsize
\begin{minipage}{0.49\textwidth}
\(
\begin{array}{@{}r@{~~}r@{~}c@{~}l@{\!}l@{}}
\textrm{Logical Formulas}
& \phi & \Coloneqq & \cdots \\
\textrm{Logical Filesystems}
  & \hat{\sigma} & \Coloneqq
  & \langle p_1\,\shorteq\,\phi_1 \cdots p_k\,\shorteq\,\phi_k \rangle \\
\textrm{Logical States}
& \lstate & \Coloneqq & \smtst{\phi}{\hat{\sigma}}
\end{array}
\)
\vskip 1em
\(
\begin{array}{r@{\,}c@{\,}l}
\smtpred{\lfs, b} & \in & \phi \\[1em]
\smtok{e} & \in & \lfs \rightarrow \kw{bool} \\
\smtok{\eskip} \lfs & \defeq & \kw{true} \\
\smtok{\eerror} \lfs & \defeq & \kw{false} \\
\smtok{\mkdir{\mkpath{p}{\str}}} \lfs &\defeq
  & \lfs(p) = \smtdir \wedge \lfs(\mkpath{p}{\str}) = \smtdne \\
\smtok{\mkfile{\mkpath{p}{\str}}{\str'}} \lfs & \defeq
  & \lfs(\mkpath{p}{\str}) \defeq  \smtdne \wedge \lfs(p) = \smtdir \\
\smtok{\rmfile{p}} \lfs & \defeq & \exists c . \lfs(p) = \smtfile{c} \wedge \\
& & \forall \str . \mkpath{p}{\str} \in \mathit{dom}(\lfs) \Rightarrow
                         \lfs(\mkpath{p}{\str}) = \smtdne \\
\smtok{\cp{p_1}{\mkpath{p_2}{\str}}} \lfs & \defeq  & \exists \str' . \lfs(p_1) = \file{\str'} \wedge \\
& & \lfs(p_2) = \dir \wedge \lfs(\mkpath{p_2}{\str}) = \dne \\
\smtok{\seq{e_1}{e_2}} \lfs & \defeq
  & \smtok{e_1}\lfs \wedge \smtok{e_2}(\smtden{e_1}\lfs) \\
\smtok{\ite{b}{e_1}{e_2}} \lfs & \defeq
  & \ite{\smtpred{\lfs, b}}{\smtok{e_1}\lfs}{\smtok{e_2}\lfs}
\end{array}
\)
\end{minipage}
\vrule
\begin{minipage}{0.48\textwidth}
\(
\begin{array}{r@{\,}c@{\,}l}
\smtden{e} & \in & \lfs \rightarrow \lfs \\
\smtden{\eskip} \lfs & \defeq  & \lfs \\
\smtden{\eerror} \lfs & \defeq  & \lfs \\
\smtden{\mkdir{\mkpath{p}{\str}}} \lfs & \defeq
  & \updfs{\lfs}{\mkpath{p}{\str}}{\dir} \\
\smtden{\mkfile{\mkpath{p}{\str}}{\str'}} \lfs & \defeq
  & \updfs{\lfs}{\mkpath{p}{\str}}{\smtfile{\str'}} \\
\smtden{\rmfile{p}} \lfs & \defeq  & \updfs{\lfs}{p}{\dne} \\
\smtden{\cp{p_1}{\mkpath{p_2}{\str}}} \lfs & \defeq
  & \updfs{\lfs}{\mkpath{p_2}{\str}}{\lfs(p_1)} \\
\smtden{\seq{e_1}{e_2}} \lfs & \defeq  & \smtden{e_2}(\smtden{e_1} \lfs) \\
\smtden{\ite{b}{e_1}{e_2}} \lfs & \defeq  &
  \ite{\smtpred{\lfs, b}}{\smtden{e_1}\lfs}{\smtden{e_2}\lfs} \\[1em]
\smt{e} & \in & \lstate \rightarrow \lstate \\
\smt{e} \smtst{b}{\lfs} & \defeq  & \smtst{b \wedge \smtok{e}\lfs}{\smtden{e}\lfs} \\[1em]
\smtg{G} & \in & \lstate \rightarrow 2^\lstate \\
\smtg{(\emptyset,E)} \lstate & \defeq  & \{ \lstate \} \\
\smtg{G} \lstate & \defeq & \bigcup \smtg{G - e}(\smt{e} \lstate) \\
& & \textrm{where $\textit{inDegree}(e) = 0$}
\end{array}
\)
\end{minipage}
\caption{Encoding \corelang as logical formulas.}
\label{encodingdefn}
\end{figure*}

\subsection{From Resource Graphs to Formulas\label{logical}}

The function $\smt{e}$ produces a collection of formulas that encode the semantics of the expression $e$ (\cref{encodingdefn}). In these formulas, two boolean variables determine whether the initial and final states are non-error states and every path is modeled by two variables that describe their initial and final state. These path-state variables are only meaningful in a non-error state. More concretely, a logical state (\lstate{}) is a record of two components: (1) \smtokf{\lstate} is a formula that is true if the current state is not the error state and (2) \smtfs{\lstate} maps paths to formulas that describe their state. We could employ McCarthy's theory of arrays~\cite{mccarthy:science-of-computation} to encode this map, but it's more efficient to encode it directly with one formula per path. To encode resource graphs $G$ as formulas, we use the function $\smtg{G}$, defined in the same figure, which maps the input logical state to a set of output logical states by evaluating each expression on the fringe with \smt{e} and recurring on the subgraph that has $e$ removed.

To prove this encoding sound and complete, we need to relate concrete states returned by the evaluator to logical states.
 This is mostly routine, but
the domain of logical filesystems has to be large enough: if a program
reads or writes to a path $p$, then there must be a formula $p \in
\textit{dom}(\smtfs{\lstate})$. For example, note that \mkdir{\texttt{/a/b}}
reads \texttt{/a} and writes \texttt{/a/b}.

\begin{lemma}[Soundness and completeness] \label{soundComplete}
\small
For all $\fs$ and $G$:

\begin{enumerate}

\item $\fs' \in\den{G}\fs$ iff there exists $\lstate \in \smtg{\smtst{\kw{true}}{\fs},e}$ such that $\lstate \vdash \smtst{\kw{true}}{\fs'}$.

\item $\errst \in \den{G}\fs$ iff there exists $\lstate \in \smtg{\smtst{\kw{true}}{\fs},e}$ such that $\smtokf{\lstate} \vdash \kw{false}$.

\end{enumerate}
\end{lemma}

\begin{figure}
\figsize
\(
\begin{array}{r@{\,}c@{\,}l}
\dom{a} & \in & 2^p \\
\dom{\pstat{p}{\isfile}} & \defeq & \{ p \} \\
\dom{\pstat{p}{\isempty}} & \defeq
  & \{ p , \mkpath{p}{\str} \}~\textrm{$\str$ is fresh} \\
\cdots \\[.25em]
\dom{e} & \in & 2^p \\
\dom{\mkdir{\mkpath{p}{\str}}} & = & \{ p, \mkpath{p}{\str} \} \\
\dom{\mkfile{\mkpath{p}{\str}}{\str'}} & = & \{\mkpath{p}{\str}, p\} \\
\dom{\rmfile{p}} & = & \{ \mkpath{p}{\str} \}~\textrm{$\str$ is fresh} \\
\dom{\cp{p_1}{\mkpath{p_2}{\str}}} & = & \{ p_1, p_2, \mkpath{p_2}{\str} \} \\
\dom{\seq{e_1}{e_2}} & = & \dom{e_1} \cup \dom{e_2} \\
\dom{\ite{a}{e_1}{e_2}} & = & \dom{a} \cup \dom{e_1} \cup \dom{e_2}
\end{array}
\)
\caption{Bounding the domain of \corelang programs.}
\label{allpathsdefn}
\end{figure}

\subsection{Checking Determinism\label{nondet}}

With resource graphs encoded as formulas, it should be straightforward to use a theorem prover to check determinism (though we have yet to address
scalability issues). Since \smtg{G} maps an input logical state to a set of output logical states, the resource graph should be deterministic, if and and only if there does not exist an input logical state that produces two different logical states. \ie, the following formula should be unsatisfiable:
\begin{mymath}
\exists \lstate_1, \lstate_2, \lstate_3 . \lstate_2 \in \smtg{G} \lstate_1 \wedge \lstate_3 \in \smtg{G} \lstate_1 \wedge \lstate_2 \ne \lstate_3
\end{mymath}
The subtlety here is that the domain of \smtg{G} to be
large enough to find a counterexample when $G$ is a non-deterministic resource graph.

To understand the issue, consider the simpler problem of checking whether
two expressions are inequivalent, $e_1 \not\deneq e_2$, which is the essence of
checking non-determinism. At first glance, it appears that expressions only
read and write to the paths that appear in it and the result of
an expression is not affected by the state of any other paths.
That is, if we have a state $\fs$ such that
$\den{e_1}\fs \not\deneq \den{e_2}\fs$ then for paths $p$ that do not appear
in either $e_1$ or $e_2$, $(\den{e_1}\fs)(p) = (\den{e_2}\fs)(p)$.
But, this equation is wrong.

The $\isempty(p)$ predicate poses a problem, since it
depends on the state of the immediate children of $p$, including those
that may not appear in the program. Consider the following inequality,
where the only difference between the programs is that one checks if the
directory is empty and the other only checks that it is a directory:
\begin{mymath}
\begin{array}{l@{\,}l}
& \ite{\pstat{\texttt{/a}}{\isempty}}{\eskip}{\eerror} \\
\not\deneq & \ite{\pstat{\texttt{/a}}{\isdir}}{\eskip}{\eerror}
\end{array}
\end{mymath}
Any input filesystem that demonstrates the inequality must have a file (or directory)
within \texttt{/a}.
However, if we construct a logical filesystem using only the paths that
appear in the program text, we will not find this counterexample.
A similar problem affects \rmfile{p}. The function in \cref{allpathsdefn}
addresses this problem by adding fresh files in directories that are
removed or tested for emptiness to avoid this bug.
We can now prove that equivalence-checking is complete.
\begin{lemma}[Completeness---equivalence]\label{equivcompleteness}
If:
\begin{itemize}
\item $\den{e_1}\fs \ne \den{e_2}\fs$  and
\item $\mathit{dom}(\lfs') = \dom{e_1} \cup \dom{e_2}$
\end{itemize}
then
$\smt{\smtst{\kw{true}}{\lfs'},e_1} \ne \smt{\smtst{\kw{true}}{\lfs'},e_2}$.
\end{lemma}
Soundness is straightforward. A model for the formula can be easily
transformed into a counterexample filesystem.
\begin{lemma}[Soundness---equivalence]\label{equivsoundness}
If:
\begin{itemize}

\item $\smt{\smtst{\kw{true}}{\lfs},e_1} \ne \smt{\smtst{\kw{true}}{\lfs},e_2}$
and

\item $\lfs \vdash \fs$
\end{itemize}
then $\den{e_1}\fs \ne \den{e_2}\fs$.
\end{lemma}

We use these lemmas to prove that that determinism checking is sound and complete.
In the theorem below, $\dom{e}$ is lifted to $\dom{G}$ in the obvious way.

\begin{theorem}(Determinism)
\label{nondetThm}
$G$ is deterministic, if and only if there exists $\lstate_1$, $\lstate_2$, and $\lstate_3$ such that $\lstate_2 \in \smtg{G} \lstate_1 \wedge \lstate_3 \in \smtg{G} \lstate_1 \wedge \lstate_2 \ne \lstate_3$ is unsatisfiable, where 
$\dom{\Sigma_1} = \dom{\Sigma_2} = \dom{\Sigma_3} = \dom{G}$.
\end{theorem}

\subsection{Commutativity and Directory Creation}\label{commutativity}

\begin{figure*}
\begin{subfigure}{0.45\textwidth}
\figsize
\(
\begin{array}{r@{\,}c@{\,}l}
\smtg{G} & \in & \lstate \rightarrow 2^\lstate \\
\smtg{(\emptyset,E)} \lstate & \defeq  & \{ \lstate \} \\
\smtg{G}\lstate & \defeq & 
\smtg{G - e}(\smt{e}\lstate)  \\
& & \textrm{where $\mathit{inDegree}(e) = 0$} \\
& & \textrm{\phantom{where} $\forall e' \in G . \neg \textit{ancestor}(e', e) \Rightarrow \seq{e'}{e} \deneq \seq{e}{e'}$ } \\
\smtg{G} \lstate & \defeq & \bigcup \smtg{G - e}(\smt{e} \lstate)  \\
& & \textrm{where $\textit{inDegree}(e) = 0$}
\end{array}
\)
\caption{Incorporating the commutativity-check.}
\label{fast_comm}
\end{subfigure}
\vrule\quad
\begin{subfigure}{0.5\textwidth}
\figsize
\(
\begin{array}{@{}r@{~~}r@{~}c@{~}l@{\!}l@{}}
\textrm{Abstract Values} & \absvcomm & \Coloneqq
  & \bot \mid R \mid W \mid D  &  \\[.25em]
\textrm{Abstract State} & \absfscomm &
  \Coloneqq & \langle p_1\!= \! \absv_1 \cdots p_k \! = \! \absv_k \rangle \\
\multicolumn{4}{l}{\bot \sqsubset R, D \sqsubset W}
\end{array}
\)
\vskip 1em
\(
\begin{array}{@{}r@{~}c@{~}l@{\!}}
\aevalcomm{e} & \in & \absfscomm \rightarrow \absfscomm \\
\aevalcomm{\idemdir{\mkpath{p}{\str}}} \absfscomm & \defeq &
\left\{\begin{array}{@{}l@{~}l}
\updfs{\absfscomm}{\mkpath{p}{\str}}{D} & \absfs(\mkpath{p}{\str}) \sqsubseteq D \\
                                        &  \textrm{and~}\absfscomm(p) = D \\
\updfs{\absfscomm}{\mkpath{p}{\str}}{W} & \textrm{otherwise}
\end{array}\right. \\
\aevalcomm{\mkdir{p}} \absfs & \defeq
  & \updfs{\absfscomm}{p}{W} \\
\aevalcomm{\mkfile{p}{\str}} \absfs & \defeq
  & \updfs{\absfscomm}{p}{W} \\
\aevalcomm{\seq{e_1}{e_2}} \absfscomm & \defeq &
 \aevalcomm{e_2}(\aevalcomm{e_1} \absfscomm) \\
 \multicolumn{3}{c}{\cdots}
\end{array}
\)
\caption{Checking commutativity.}
\label{commcheck}
\end{subfigure}
\caption{Commutativity checks eliminate the number of permutations that need to be generated.}
\end{figure*}

Modeling all valid permutations of resources can produce formulas that are intractably large. For example, suppose a resource graph $G$ has exactly two nodes $a$ and $b$ that do not have any ancestors. The naive approach considers evaluating either node first and then recurs on the two subgraphs $G - a$ and $G - b$. When the sub-graphs also have several nodes without any ancestors, the size of the generated formula grows intractably large. A significantly better approach is to use a fast, syntactic commutativity check to rule out permutations that don't need to be explored, similar to partial-order reduction. Note that it is not sufficient to check that $a$ and $b$ commute. For example, \seq{b}{\seq{a}{c}} and \seq{b}{\seq{c}{a}} are valid permutations in the following graph:
\begin{center}
\begin{tikzpicture}
\node at (0,1) {$a$};
\node(2) at (1,1) {$b$};
\node(3) at (2,1) {$c$};
\path (2) edge[right,->] (3);
\end{tikzpicture}
\end{center}
We can only conclude that they are equivalent if we know that $a$ commutes with both $b$ and $c$. Therefore, to avoid recurring on both $G - a$ and $G - b$, we need to prove that $a$ (or $b$) commute with all nodes that are not ancestors of $a$, as shown in \cref{fast_comm}.

Next, we need a fast, syntactic commutativity check, which should be straightforward to do for \corelang. Surprisingly, the natural approach does not work. A typical commutativity check works as follows: to check if $e_1$ and $e_2$
commute, calculate the set of locations that each reads and writes. If
the expressions don't have any overlapping writes and $e_1$ does not read
any locations that $e_2$ writes (and vice versa), then they do commute.
If not, they may or may not commute and we need to semantically check both
orderings.

This approach is not effective for Puppet, due to the semantics of packages.
Typical packages install files to shared directories (\eg\, \texttt{/usr/bin},
\texttt{/etc}, and so on) and will create these directories if necessary. Therefore, the conventional approach cannot prove that packages
commute. Manifests that installs several packages typically do not specify any
dependencies between them, so this issue arises frequently.

To address this issue, we use an abstract interpretation that
maps each path $p$ to the abstract values $\bot$, $R$, $W$, and $D$ (\cref{commcheck}). These values indicate that the expression either
does not affect $p$ ($\bot$), reads from $p$ ($R$), writes to $p$ ($W$), or
ensures that $p$ is a directory ($D$).
A \mkdir{p} expression that doesn't first check if $p$ already exists
is simply a write ($W$). Only a guarded \mkdir{p} can ensure $p$ is a
directory, such as these expressions:
\begin{mymath}
\begin{array}{ll}
& \idemdir{p} \\
\deneq & \ite{\pstat{p}{\dne}}{\mkdir{p}}{\ite{\pstat{p}{\isfile}}{\eerror}{\eskip}}
\end{array}
\end{mymath}
In addition, the analysis ensures that expressions create directory
trees in a reasonable order. For example, an expression
that creates \texttt{/a} before \texttt{/a/b} is not equivalent
to an expression that tries to create \texttt{/a/b} before \texttt{/a}.
However, two expressions that create sibling directories do commute.
To ensure that these properties hold, we map \mkpath{p}{\str}
to $D$, only if $p$ is already mapped to $D$.

We can use the result of this analysis to check that expressions
commute, even if they create overlapping directory trees.
{
\newcommand{\lhs}{\aevalcomm{e_1}\bot(p)}
\newcommand{\rhs}{\aevalcomm{e_2}\bot(p)}
\begin{lemma} \label{commLem} For all $e_1$ and $e_2$, if:
{\small
\begin{enumerate}

  \item $\{ p \mid \lhs = R \} \cap \{ p \mid \rhs = W \} = \emptyset$,
  \item $\{ p \mid \lhs = W \} \cap \{ p \mid \rhs = R \} = \emptyset$,
  \item $\{ p \mid \lhs = D \} \cap \{ p \mid \rhs\in\{R,W\} \} = \emptyset$, and
  \item $\{ p \mid \lhs\in\{R,W\} \} \cap \{ p \mid \rhs = D \} = \emptyset$

\end{enumerate}}
then $\seq{e_1}{e_2} \deneq \seq{e_2}{e_1}$.
\end{lemma}
}

\subsection{Pruning Files from Resources\label{pruningsec}}

The syntactic commutativity check mentioned above eliminates the need to explore
different permutations of resources that are obviously equivalent to each other.
However, even a single permutation that installs several large resources can
make formulas needlessly large. For example, suppose a manifest installs a large
package (\eg, \texttt{git}, which has over 500 files) and then doesn't read or
write to any of the files that the package creates. Intuitively, we should
be able to completely \emph{eliminate} resources that are not observed by
the rest of the manifest.

However, there are situations where resources must interfere.
It is quite common for a manifest to update a default configuration file created
by a package. For example, the manifest in \cref{apache2} installs the Apache 
web server and supplies a site-specific configuration file that should overwrite
the default configuration. Even in this situation, the manifest does not update most of the other 200+ files
that the Apache package creates. Intuitively, we should be able to \emph{shrink}
resources so that we don't have to track the state of files that cannot affect the outcome of
the determinism-check.

In this section, we formalize these two observations using two simple analyses.

\paragraph{Eliminating Resources.}
Notice that a determinism-check is essentially a conjunction of equivalence-checks
between all valid permutations of resources. For example, the following resource graph has eight valid permutations of the four resources shown:
\begin{center}
\begin{tikzpicture}
\node(1) at (0,1) {$a$};
\node(2) at (1,1) {$c$};
\node(3) at (2,1) {$b$};
\node(4) at (3,1) {$d$};
\path (1) edge[right,->] (2);
\path (3) edge[right,->] (2);
\end{tikzpicture}
\end{center}
A naive determinism-check would generate all permutations and verify that they are equivalent:
\begin{mymath}
\begin{array}{l@{\,}l@{\,}l@{\,}l@{\,}l@{\,}l@{\,}l@{\,}l}
         & \seq{a}{\seq{b}{\seq{c}{d}}}
& \deneq & \seq{a}{\seq{b}{\seq{d}{c}}} 
& \deneq & \seq{a}{\seq{d}{\seq{b}{c}}}
& \deneq & \seq{b}{\seq{a}{\seq{c}{d}}} \\
\deneq & \seq{b}{\seq{a}{\seq{d}{c}}} 
& \deneq & \seq{b}{\seq{d}{\seq{a}{c}}} 
& \deneq & \seq{d}{\seq{a}{\seq{b}{c}}}
& \deneq & \seq{d}{\seq{b}{\seq{a}{c}}}
\end{array}
\end{mymath}
However, suppose we use our commutativity check to determine that $c$ and $d$ commute. We could then rewrite all the permutations that end with \seq{c}{d} to instead end with \seq{d}{c}, which gives us a series of permutations that all end in $c$. In general, 
$\seq{e_1}{e} \deneq \seq{e_2}{e}$, if and only if $e_1 \deneq e_2$. Therefore, we can completely eliminate $c$ without changing the result of the equivalence check.

In general, if a resource commutes with all other resources that may be evaluated after it in the resource graph, then that resource can be eliminated without affecting the result of the determinism-check. Moreover, eliminating one resource often allows their parents to be eliminated. For example, suppose that $b$ commutes with $a$ and $d$. Eliminating $c$, as discussed above, allows us to then eliminate $b$ by the same argument. However, trying to eliminate $b$ first would fail, since it does not commute with $c$, which may-succeed $b$.
In practice, a true dependency $a \rightarrow b$ indicates that $b$ truly depends on the effects of $a$ and thus the two resources do not commute. In our experience, we've found it most effective to eliminate resources by starting with resources at the fringe of the dependency graph that are not required by any other resources. 

\paragraph{Shrinking Resources.}
There are several cases where large resources cannot be completely
eliminated. However, they can be shrunk as follows.
In general, if a resource writes to a path $p$ such that (1) other resources
in the manifest do not observe the state of $p$ and (2) other resources in the
manifest do not affect the state of $p$ then we can eliminate
writes to $p$ without changing whether the manifest is deterministic or not.
Moreover, the encoding of \corelang{} programs as formulas can then exploit the fact
that $p$ is read-only and use a single variable to represent the state of $p$, 
instead of using new variables for each state. This can dramatically reduce
the number of variables needed to encode the program.

Consider the problem of shrinking two expressions
$e_1$ and $e_2$ to $e_1'$ and $e_2'$, such that $e_1 \deneq e_2$ if and
only if $e_1' \deneq e_2'$. If both expressions leave a path $p$ in the same state, it should be
possible to shrink both expressions by removing their writes to $p$. However, to implement idempotent
operations, resources tend to have a complex series of reads and writes
(\cref{compilingresources}). Nevertheless, a resource that writes to $p$
typically ensures that $p$ is either placed in a definite state or signals an error if it cannot do so. We say that these resources make \emph{definitive writes} to $p$.
Therefore, if both expressions make the same definitive write to $p$, then we can eliminate writes to $p$.

We detect definitive writes using the abstract interpretation sketched in \cref{defnwrites}, which produces an abstract
heap, \absfs{} that maps paths $p$ to abstract values that characterize
the effect of an expression on $p$ over all input states:
\begin{itemize}

  \item If $\absfs(p) = \smtdir$, the expression ensures that $p$ is a
   directory (or signals an error).

  \item If $\absfs(p) = \smtfile{\str}$, the expression ensures that $p$
  is a file with contents \str (or signals an error).

  \item If $\absfs(p) = \smtdne$, the expression ensures that $p$ does not
  exist (or signals an error).

  \item If $\absfs(p) = \bot$, the expression does not read or write $p$.

  \item If $\absfs(p) = \top$, the expression has an indeterminate effect
  on $p$.

\end{itemize}

\begin{lemma}
If $(\aeval{e}\bot)(p) \sqsubset \top$ then for all states $\fs_1$ and
$\fs_2$, $(\den{e}(\fs_1))(p) = (\den{e}(\fs_2))(p)$.
\end{lemma}

\begin{figure*}
\figsize
\begin{subfigure}{0.66\textwidth}
  \(
  \begin{array}{r@{\,}c@{\,}l}
  \pruneRec{}{\cdot}{} & \in & e \times p \times \fs \rightarrow e \times \fs \\
  \pruneRec{p}{\eskip}{\fs} & = & (\eskip, \fs) \\
  \pruneRec{p}{\eerror}{\fs} & = & (\eerror, \fs) \\
  \pruneRec{p}{\mkdir{p}}{\fs} & =
    & (\eerror, \fs)~\textrm{if $\fs(p) = \dir$ or $\fs(p) = \file{\str}$} \\
  \pruneRec{\mkpath{p}{\str}}{\mkdir{\mkpath{p}{\str}}}{\fs} & = &
  (\ite{\dne(\mkpath{p}{str}) \wedge \isdir(p)}{\eskip}{\eerror},
  \updfs{\fs}{\mkpath{p}{str}}{\dir}) \\
  \pruneRec{p}{\mkdir{p'}}{\fs} & = & (\mkdir{p'},\fs)~\textrm{if $p \ne p'$} \\
  \pruneRec{p}{\mkfile{p}{\str}}{\fs} & =
    & (\eerror, \fs)~\textrm{if $\fs(p) = \dir$ or $\fs(p) = \file{\str'}$} \\
  \pruneRec{\mkpath{p}{\str}}{\mkfile{\mkpath{p}{\str}}{\str'}}{\fs} & = &
  (\ite{\dne(\mkpath{p}{str}) \wedge \isdir(p)}{\eskip}{\eerror},
  \updfs{\fs}{\mkpath{p}{str}}{\file{\str'}}) \\
  \pruneRec{p}{\mkfile{p'}{\str}}{\fs} & = & (\mkfile{p'}{\str},\fs)~\textrm{if $p \ne p'$} \\
  \pruneRec{p}{\rmfile{p}}{\fs} & = & (\eerror, \fs)~\textrm{if $p \not\in \dom{\fs}$ or $\exists \str . \mkpath{p}{\str} \in \dom{\fs}$} \\
  \pruneRec{p}{\rmfile{p}}{\fs} & = &
      (\ite{\por{\isfile(p)}{\isempty(p)}}{\eskip}{\eerror}, \fs - p) \\
  \pruneRec{p}{\rmfile{p'}}{\fs} & = & (\rmfile{p'}, \fs)~\textrm{if $p \ne p'$} \\
  \multicolumn{3}{c}{\cdots} \\
  \pruneRec{p}{\ite{a}{e_1}{e_2}}{\fs} & = & (e_1,\fs)~\textrm{if $\den{a}\fs = \ptrue$} \\
  \multicolumn{3}{c}{\cdots} \\[.5em]
\textit{prune} & \in & p \times e \rightarrow e \\
\prune{p}{e} & = & p'~\textrm{where $(p', \sigma) = \pruneRec{p}{e}{\cdot}$}
  \end{array}
  \)
\caption{Pruning definitive writes.}
\label{pruning}
\end{subfigure}
\vrule~~
\begin{subfigure}{0.32\textwidth}
\(
\begin{array}{@{}r@{~~}r@{~}c@{~}l@{\!}l@{}}
\textrm{Abs. Values} & \absv & \Coloneqq
  & \bot \mid \top \mid \smtdir \mid \smtfile{\str} \mid \smtdne &  \\[.25em]
\textrm{Abs. State} & \absfs &
  \Coloneqq & \langle p_1\!= \! \absv_1 \cdots p_k \! = \! \absv_k \rangle \\[1em]
\multicolumn{4}{c}{\bot \sqsubset \smtdir, \smtfile{\textit{str}}, \smtdne \sqsubset \top}\end{array}
\)
\vskip 1em
\(
\begin{array}{@{}r@{~}c@{~}l@{\!}}
\aeval{e} & \in & \absfs \rightarrow \absfs \\
\aeval{\eskip} \absfs & = & \absfs \\
\aeval{\eerror} \absfs & = & \absfs \\
\aeval{\mkdir{p}} \absfs & = & \updfs{\absfs}{p}{\smtdir} \\
\aeval{\mkfile{p}{\str}} \absfs & = & \updfs{\absfs}{p}{\smtfile{\str}} \\
\aeval{\rmfile{p}} \absfs & = & \updfs{\absfs}{p}{\smtdne} \\
\aeval{\ite{a}{e_1}{e_2}} \absfs & = & \aeval{e_1} \absfs \sqcap \aeval{e_2} \absfs \\
\aeval{\seq{e_1}{e_2}} \absfs & = & \aeval{e_2}(\aeval{e_1} \absfs)
\end{array}
\)
\caption{Detecting definitive writes.}
\label{defnwrites}
\end{subfigure}
\caption{Shrinking resources.}
\end{figure*}

If the abstract interpretation determines that  $e_1$ and $e_2$ set
a path $p$ to the same definite value, we should be able to prune writes
to $p$ from both expressions. However, consider the two equivalent
expressions below:
\begin{mymath}
  \seq{\mkdir{\mkpath{p}{\str}}}
      {\ite{\isdir(\mkpath{p}{\str})}{\eskip}{\eerror}}  \equiv
  \mkdir{\mkpath{p}{\str}}
\end{mymath}
The expressions on either side ensure that \mkpath{p}{\str} is a directory. 
However, if we naively replace \mkdir{\mkpath{p}{\str}} with \eskip, we get
the following wrong result:
\begin{mymath}
\seq{\eskip}{\ite{\isdir(\mkpath{p}{\str})}{\eskip}{\eerror}} \neq
  \eskip
\end{mymath}
To correctly eliminate writes to $p$, we need to also transform
expressions that read from $p$ to account for the effect that
the write would have had. In our example, the test 
\isdir(\mkpath{p}{\str}) will always be true, since it follows 
\mkdir{\mkpath{p}{\str}}.
The insight is that when  writes to $p$ are eliminated, we need to transform
all expressions that subsequently read or write to $p$. In our example, we
need to transform \isdir(\mkpath{p}{\str}) to \ptrue.

The pruning function, $\prune{p}{e}$, eliminates writes to $p$ by preserving
reads in this manner (\cref{pruning}). The function correctly handles
programs where a write to $p$ is followed by other reads and writes to $p$
by partial evaluation.

The following lemma states that the same definitive write from $e_1$
and $e_2$ doesn't change their (in-)equivalence.
\begin{lemma} \label{pruningLem}
If $(\aeval{e_1}\bot)(p) = (\aeval{e_2}\bot)(p) = \absv$ and $\absv \sqsubset \top$ then $e_1 \deneq e_2$ if and only if $\prune{p}{e_1} \deneq \prune{p}{e_2}$.
\end{lemma}

Although pruning eliminates writes to $p$, it does not eliminate reads from $p$.
However, eliminating writes ensures that $p$ is a read-only path. When we
encode the expression as a logical formula, the encoding can optimize for
read-only paths by using a single variable to represent the initial state of the
path, which then remains unchanged.

\paragraph{Pruning for determinism checking.}
Since a determinism check encodes equalities between all permutations
of resources, we could also
apply the abstract interpretation to all permutations, but this would be
intractable. Instead, we apply the abstract interpretation to each resource
in isolation to find paths that are definitively written by exactly one
resource and only prune these paths. This conservative approach works
well in practice.

\subsection{Summary}

These are the three major techniques that \system uses to make
determinism-checking scale.  We've also outlined how each step preserves (in-)equivalences,
so the approach is sound and complete.

\paragraph{Other approaches.}
We have tried two other techniques for checking determinism that
are less effective than the methodology discussed in this section.
\begin{enumerate}

\item We developed a dynamic analysis that simply installed
resources in different valid permutations within independent 
Docker containers. The Docker API makes it easy to see how a container
has updated its filesystem. However, installing resources takes time
and it took our prototype several hours to verify small manifests
with less than ten resources. (We fully utilized a four-core machine
with 16 GB RAM.) In contrast, our static analysis
checks determinism in seconds.

\item Instead of using an SMT solver, we tried to encode \corelang
programs as binary-decision diagrams (BDDs) by exploiting the natural
hierarchy of paths to pick a good variable order. (\eg, 
$\texttt{a} < \texttt{a/b}$.) In our experience, the SMT solver
was faster and significantly easier to use. For example, properties such as 
``all paths must be distinct'' are very easy to express using
\texttt{distinct} constraints in an SMT solver.

\end{enumerate}

\section{Beyond Determinism}
\label{idempotence}

After we've checked that a manifest is deterministic, we can treat it as an
expression rather than a resource graph: we can pick any valid
ordering of the resources (determinism ensures that they are all equivalent)
and sequence them to form a single expression $e$. We emphasize that
while resource graphs denote relations, \corelang{} expressions denote functions.
This lets \system check several properties quickly and easily.

\paragraph{Invariants.}
We've seen that Puppet is actually very imperative. A manifest that
declares a \texttt{file} resource may overwrite it using some other
resource, which is typically undesirable. \system checks for this issue
using the following formula, which is unsatisfiable if $e$ ensures
that $p$ is always a file with content \str:
\begin{mymath}
\exists \lfs . \smtok{e}\lfs \wedge \smtden{e}\lfs(p) \ne \smtfile{\str}
\end{mymath}
It is easy to imagine checks for several other invariants.

\paragraph{Idempotence.}
We discussed in \cref{overview} that idempotence is a critical property of
Puppet manifests. To test if a manifest is idempotent, we simply check
if $e \deneq \seq{e}{e}$ holds.

We emphasize that these checks are efficient because they do not have
to consider all permutations of resources. Moreover, these simple checks would be unsound if applied to non-deterministic manifests.

\section{Evaluation\label{evaluation}}

\system is implemented in Scala and uses the Z3
Theorem Prover~\cite{demoura:z3} as its SMT solver. The majority of the
codebase is the frontend that turns manifests into \corelang expressions. To
model packages, \system needs to query an OS package manager.
For portability, we've built a web service for \system that can query the
package manager for several operating systems.
The service returns the package listing in a standardized format and stores
the result in a database to speedup subsequent queries. Our current deployed
service has Ubuntu and CentOS running in containers and it is easy to add support for
other operating systems.

Note that the times reported in this section do not include the time required to fetch package listings. The package querying tools (\texttt{apt-cache} and \texttt{repoquery}) can take several seconds to run, which is why our web service caches their results.

\paragraph{Third-party benchmarks.}
We benchmark the determinism checker on a suite of 13 Puppet configurations gleaned from
GitHub and Puppet Forge. We specifically chose benchmarks that did not use
\verb|exec| resources as detailed further in \cref{limitations}. We manually verified
that six of them have determinism bugs and that seven do not. For each non-deterministic
program, we developed a fix and verified that \system reports that it is deterministic
and idempotent. We repeat all timing experiments ten times and report the average. We
perform all experiments on a quad-core 3.5 GHz Intel Core i5 with 8GB RAM.

\begin{figure*}
\centering
\begin{subfigure}{.3\textwidth}
\begin{tikzpicture}
\node{\pgfimage[width=0.9\textwidth]{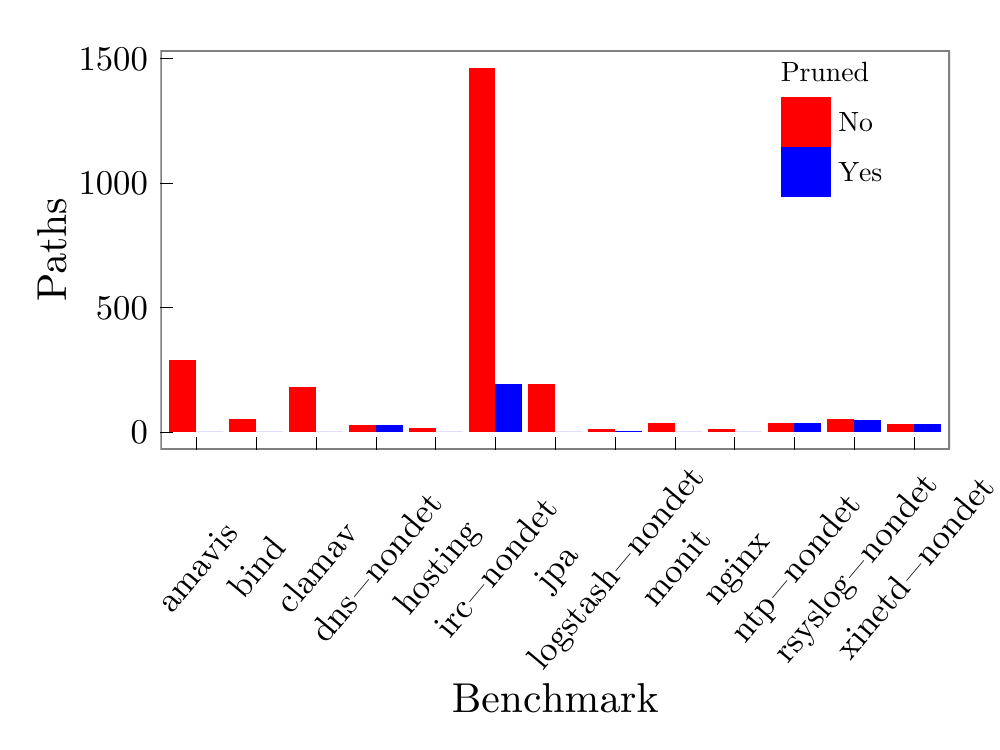}};
\end{tikzpicture}
\caption{Paths per state.}\label{pathbench}
\end{subfigure}
\begin{subfigure}{.3\textwidth}
\begin{tikzpicture}
\node{\pgfimage[width=0.9\textwidth]{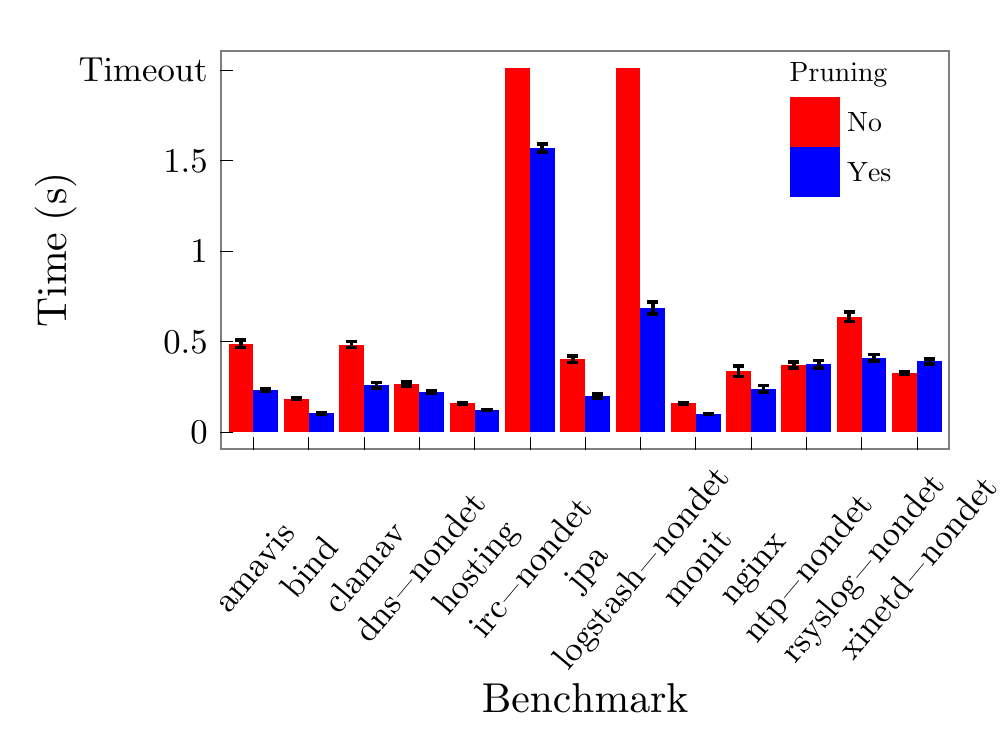}};
\end{tikzpicture}
\caption{Shrinking and eliminating resources.}\label{deterbench}
\end{subfigure}
\begin{subfigure}{.3\textwidth}
\begin{tikzpicture}
\node{\pgfimage[width=0.9\textwidth]{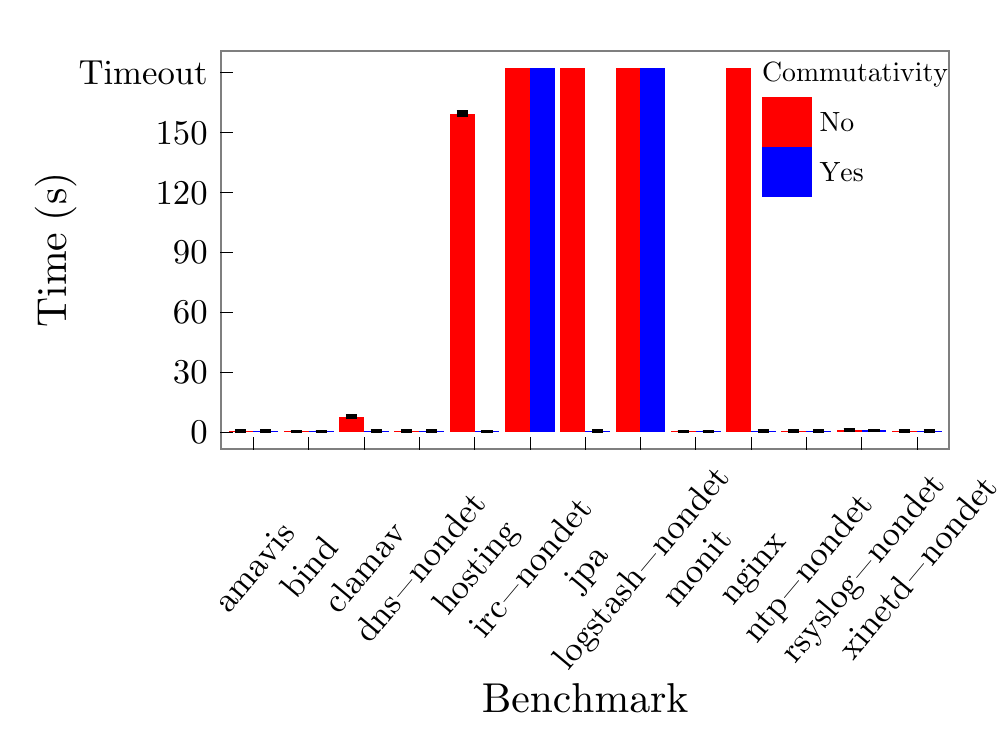}};
\end{tikzpicture}
\caption{Checking commutativity.}\label{commbench}
\end{subfigure}
\caption{Benchmarking determinacy analysis.}
\label{deter-major-figure}
\end{figure*}

\begin{figure}
\centering
\begin{tikzpicture}
\node{\pgfimage[width=0.26\textwidth]{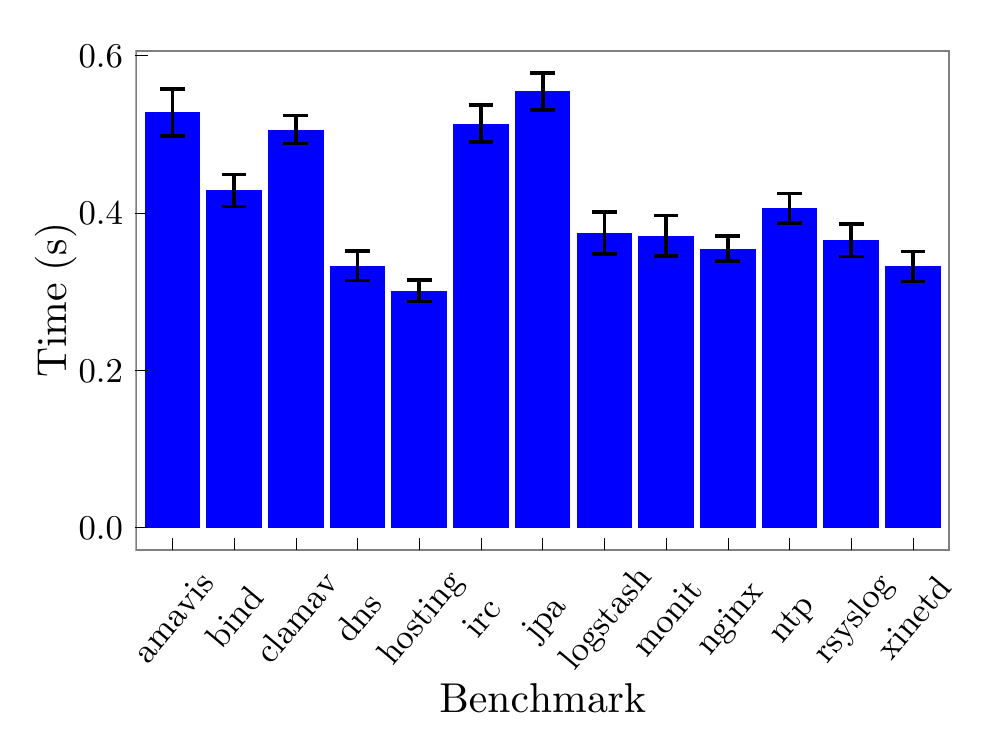}};
\end{tikzpicture}
\caption{Benchmarking idempotence checking.}\label{idembench}
\end{figure}

\Cref{deter-major-figure} shows the effect of pruning on \system{}'s determinacy analysis. 
In the figure, the
non-deterministic manifests are marked \emph{-nondet}.
Without commutativity checking, four benchmarks do not complete in over ten minutes and one takes over two minutes to run (\cref{commbench}).
Even with commutativity-checking, two benchmarks timeout after ten minutes. However, when commutativity-checking is coupled with file pruning (\cref{pruningsec}), all benchmarks complete in less than two seconds (\cref{deterbench}). 
\Cref{pathbench} shows the number of files in each manifest, with and without pruning. (Note that commutativity-checking does not affect the number of files.) As expected, the runtime of benchmarks corresponds to the number of files that need to be modeled. 

\Cref{idembench} reports the time required by the idempotence check is less
than one second on all benchmarks. In practice, the idempotence check would
be preceded by a determinism check, which typically takes more time to complete.

\begin{figure}
\centering
\begin{tikzpicture}
\node{\pgfimage[width=0.26\textwidth]{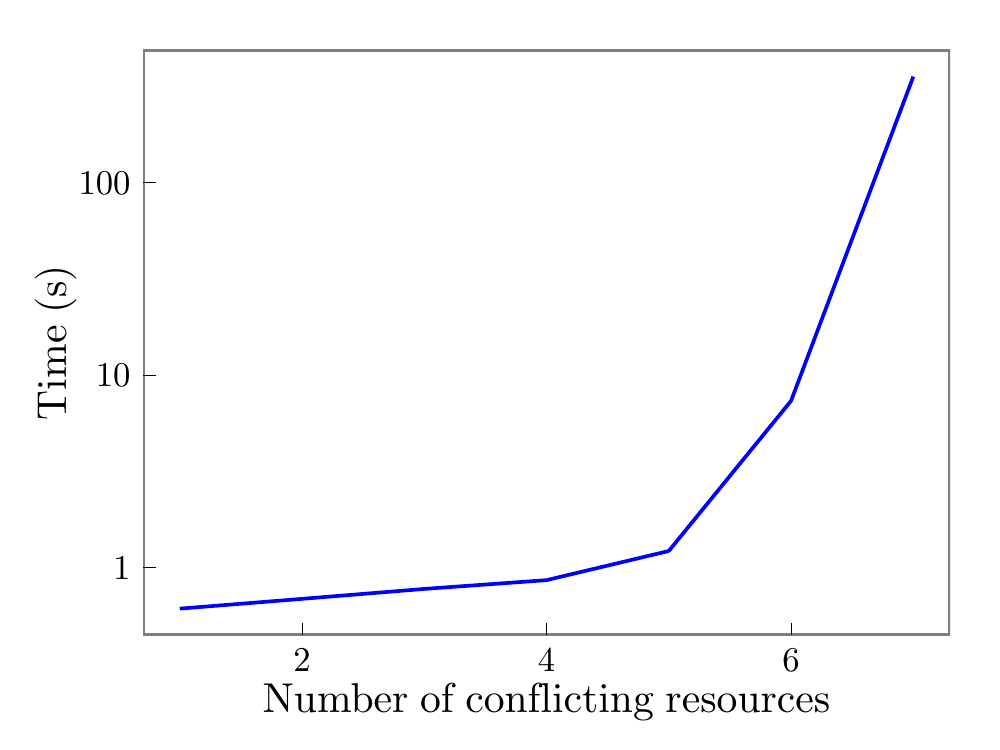}};
\end{tikzpicture}
\caption{Scalability with $n$ interfering resources.}\label{scalefail}
\end{figure}

\paragraph{Synthetic benchmarks.}
The benchmarks above suggest that commutativity checking and pruning are effective in practice. 
However, it is quite straightforward to construct an artificial scenario where the commutativity check is ineffective.
The natural way to construct this benchmark is to have $n$ unordered file-resources that write to the same path. This renders the commutativity-check useless, so \system{} is forced to explore all $n!$ paths through the resource graph. Moreover, the file cannot be pruned either. \Cref{scalefail} shows the running time grows non-linearly with $n$. In fact, even with $n = 6$, the running time exceeded two minutes.

Although the simple benchmark described above can be constructed using \corelang{}, it is not a valid Puppet manifest, since Puppet does not allow multiple file-resources to affect the same path. A working alternative is to find $n$ conflicting packages that
all create the same file $p$ and try to install all of them simultaneously. Even in this scenario, 
\system{} can determine that the manifest is non-deterministic relatively quickly (\ie, that
the formula is satisfiable). However, we can force the manifest to be 
deterministic by using a single file resource that updates the contents of $p$ after all
the packages are installed:
\begin{puppetcode}
# All packages create a file /a
package{'A-1': before => File['/a'] }	
package{'A-2': before => File['/a'] }
package{'A-3': before => File['/a'] }
...
file{'/a': content => 'x' }
\end{puppetcode}
The final  file-resource makes the manifest deterministic, which forces the
solver to construct a proof of unsatisfiability instead of terminating early with a satisfying
assignment. We believe that this kind of scenario is very unlikely to arise in practice.

\paragraph{Bugs found.}
\system found determinism bugs in six benchmarks (including a previously undiscovered
bug).  The bugs are of the kind we described in \cref{overview}. Specifically, several
benchmarks omitted a necessary dependency between a package and a configuration file.
In addition, one benchmark omitted a dependency between a user account and SSH keys for
the user. Broadly speaking, resource-types such as files and packages have a
well-understood semantics, but users may not understand their interactions.

\section{Related Work\label{related}}

\paragraph{Other system configuration languages.}
Several system configuration
languages have been developed over decades of research, many of which are surveyed
by Delaet and Joosen~\cite{delaet:sysconfig}. To the best of our
knowledge, the kind of verification tools we have developed for
Puppet have not been developed for these languages. Instead, we
highlight how several languages differ from Puppet and consider what it would take
to adapt our approach for them.

Hagemark and Zadeck's Site tool~\cite{hagemark:site}
 has a DSL that is closely related to Puppet. A Sitefile describes
bits of configurations in ``classes'' that can be composed in
several ways. Site traverses these classes in topological order
and  can also suffer missing dependencies, which our techniques  detect.

LCFG~\cite{anderson:lcfg} provides built-in components for configuring common applications.
However, while new LCFG components have to be authored in Perl, Puppet
encourages average users to build their own abstractions using the Puppet DSL.
An inter-component dependency in LCFG requires coordination between the
configuration file and Perl code (using ``context variables'').
\system leverages Puppet's high-level DSL which makes all dependencies
manifest. Building similar tools for LCFG would be difficult due to Perl.

Anderson and Herry~\cite{anderson:smartfrog} develop a denotational semantics for the SmartFrog
configuration language that faithfully models its non-deterministic
semantics.  They show that their model helps resolve several
implementation issues, though ordering issues remain. They argue that system configuration languages need
formal models and warn that popular languages gain features
faster than formal models can be developed. Our work shows that it is possible
to model and analyze a significant fraction of a large system configuration
language, but we don't disagree with their conclusions.

Engage~\cite{fischer:engage} is a system for deploying and configuring
distributed applications that can specify
complex, inter-machine dependencies, where values computed by one resource
at runtime can be used as inputs to another resource on a different machine.
Puppet is more limited and does not support orchestration.
To manage the life-cycle of a resource, Engage users have to write drivers in Python. Although the Engage type-checker ensures that resources are composed
correctly, it assumes that these Python drivers are error-free. In contrast,
the Puppet DSL performs operations similar to Engage drivers and
our tools can check this code.

NixOS~\cite{dolstra:nixos} takes a radically different approach to package and
configuration management than a typical Linux distribution. NixOS places every
package and configuration in a unique location (determined during configuration)
and ensures that they are immutable. This design forces NixOS policies to make
all dependencies explicit. Puppet bring some of the advantages of NixOS to
traditional operating systems and Linux distributions, but our paper shows that
it doesn't provide the same guarantees of NixOS. Instead of proposing a
radical, new architecture, we show that program verification techniques can be
employed to provide strong guarantees for Puppet configurations.

Tucker and Krishnamurthi~\cite{tucker:config} argue
that Racket's unit system could be adapted to build a better package manager.
The benefits of their design are similar to the benefits of NixOS (discussed
above).

\paragraph{Testing and verification of configurations.}
\textsc{CloudMake} is a cloud-based build system in use at
Microsoft that has important features such as artifact caching, parallel builds,
etc. \textsc{CloudMake} commands make all inputs and outputs explicit.
Christakis, et al.~\cite{christakis:cloudmake} have a mechanized proof
that \textsc{CloudMake} scripts are race-free, which justifies parallel builds.
Our paper shows that it's not possible to prove such a theorem for
all Puppet configurations. Instead, \system verifies that
individual manifests are deterministic.

Hummer et al.~\cite{hummer:chef-idem}
systematically test Chef configurations and find that several configurations
are not idempotent. Their test-based approach cannot ensure complete
coverage and can take several days. By contrast, we use static analysis to prove
determinacy and idempotency, which would be more difficult to do for Chef as it is a Ruby-embedded DSL.

Although Puppet uses native package managers to implement package
resources, Puppet doesn't leverage the rich information that packages
provide, such as their direct dependencies and conflicts, which leads
to the kind of errors described in \cref{overview}.
It should be possible to leverage package metadata to build more
useful verification tools, perhaps using the SAT-based encoding
of Opium~\cite{tucker:opium}. Unlike \verb|apt-get|,
Opium's algorithm for calculating installation/uninstallation
is complete for a given distribution.
The analogous problem for Puppet would be to calculate the installation
profile for a resource, given a universe of resources, such as modules on
Puppet Forge. To do so, one would need to
calculate and verify dependencies. \system does the latter
and could be augmented to do the former.

\system uses a  straightforward model of the filesystem, partly because
Puppet's model hides many platform-specific filesystem details for
portability (\eg, Puppet doesn't support hard links). Others have
developed filesystem models that are much richer than ours (\eg, \cite{ridge:sibylfs,arkoudas:fs,morgan:unixfs}). The program
logic of Gardner et al.~\cite{gardner:posix} is particularly interesting
because it enables modular reasoning about filesystem-manipulating programs. In
contrast, the verification techniques in our paper are not modular because
we support  Puppet features that have global effects on
the resource graph. If these features were ignored, a modular analysis would
be attractive.

Cloud services such as Microsoft Azure contain large configurations with many 
components in various representations (\eg YAML, XML, INI, etc.). 
ConfValley~\cite{huang:confvalley} unifies these configurations into a
single representation and validates them with respect to user-written 
predicates about the configuration. The predicates may describe desirable 
properties for a particular cloud service configuration such as ensuring that a 
particular variable has the correct type or a certain file has the appropriate 
permissions. Rehearsal verifies two specific properties about the effects of a Puppet configuration on a machine, rather than properties of the 
configuration itself. We consider every possible input and execution path in 
order to prove or disprove idempotence and determinism. A ConfValley-style
verification of Puppet would involve writing predicates about the structure
of the resource graph, which should be straightforward to do with our tools.

\paragraph{Determinacy checkers.}
In the past few years, several tools have been developed that use
static~\cite{kawaguchi:dpar-liquid,vechev:dice,bocchino:dpj}
and dynamic~\cite{burnim:determinism,sadowski:singletrack} techniques 
to check that multi-threaded programs are
deterministic. \system is a static determinacy checker for Puppet and
leverages an SMT solver, thus is most closely related to Liquid
Effects~\cite{kawaguchi:dpar-liquid}.
Liquid Effects establishes determinism by showing that concurrent 
effects are disjoint, but there are common examples of deterministic 
Puppet programs that do not have disjoint effects. Instead, \system has 
a commutativity check that accounts a pattern of false sharing that is 
common to Puppet (\cref{commutativity}). \system and Liquid Effects 
address determinism in two very different domains. Liquid Effects 
proves determinism for multi-threaded C programs with pointers, 
aliasing, and functions that are tackled in a modular way with types. 
In contrast, Puppet manifests have no aliasing, loops, or procedures. 
Since our problem is simpler, we are able to build a scalable, sound, 
and complete determinacy checker that requires no annotations by the 
programmer.

When \system reports that a Puppet manifest is \emph{deterministic},
the manifest may still yield different outputs for different inputs.
\ie, \system only verifies that a manifest maps each input state to
a single output state. In contrast, Andreasen and 
M{\o}ller~\cite{andreasen:tajs},  have developed techniques to infer
that program expressions \emph{determinate}, \ie, that an expression
produces the same value in all executions. They exploit determinate
expressions to improve the precision of their JavaScript Type Analyzer. 
In contrast, Puppet expressions are always determinate, but Puppet
manifests can be non-deterministic.

\paragraph{Alternate uses of configuration management.} Finally, we note that  configuration management is an overloaded term in the literature. This paper addresses an issue that arises in software configuration and deployment. However, the term configuration management is also used to refer to version-control systems (\eg, CVS and Git) and to application configuration~\cite{tang:facebook-config,sherman:acms}, which is not the subject of this work.

\section{Limitations\label{limitations}}

The primary limitation of this work is that Puppet manifests support 
embedded shell scripts (using the \texttt{exec} resource type). Shell 
scripts are often an anti-pattern, but they are indispensable for certain 
tasks. For example, they are often used to setup software that has not
made its way into sanctioned software repositories. The main challenge
with shell scripts is that they can have arbitrary effects on the
filesystem, unlike the other resource-types that have a clearer semantics
and lend themselves to formal models.

Another limitation of our work is that our analyses rely on models of
system resources, which can be inaccurate. For example, to model packages,
we need to know the files that a package creates. At present, we assume
that packages only create the files returned by \texttt{apt-file} (on 
Debian) and \texttt{repoquery} (on Red Hat). However, many packages
use ``post-install scripts'' to create additional files, which our
approach will miss. Therefore, although our algorithms are sound and 
complete with respect to our model of system resources, our models
have known limitations. A more precise alternative would be to actually 
install packages in a sandboxed environment and check what files get 
written to disk.

Finally, as suggested above, our analysis is platform-dependent.
In fact, the choice of operating system determines how packages are
modeled. Although Puppet has several platform-neutral features, it also
exposes the platform name and version as program variables that a manifest can use to specialize for a particular platform.
\system{} takes the platform name as a command-line flag and so a manifest can be re-verified for several platforms.
However, it would be more useful to check that a manifest has similar effects on different platforms.

\section{Conclusion\label{conclusion}}
This paper presents \system, the first verification tool for Puppet, a popular
system configuration language. Specifically, \system checks that \texttt{exec}-free
Puppet manifests are deterministic and idempotent, which are both fundamental
properties of correct Puppet manifests. To build \system, we developed a simple
semantics for Puppet that we hope will be useful to other researchers.  We
believe that our approach to modeling Puppet will enable several other tools,
\eg, manifest repair and synthesis, and security auditing.

\acks

We thank the PLDI'16 reviewers, our shepherd
Manu Sridharan, Daniel Barowy, Emery Berger, Shriram Krishnamurthi, Robert 
Powers, John Vilk, and Jean Yang for their thoughtful feedback and 
suggestions. We thank Joseph Collard and Nimish Gupta for their work on a 
preliminary version of \system. This work is supported by the U.S. 
National Science Foundation under grants CNS-1413985 and CCF-1408745
and by a Google Research Award.

\bibliographystyle{plainnat}
\bibliography{../bib/venues-long,../bib/papers}

\end{document}